RESEARCH ARTICLE                                                                                   OPEN ACCESS

# Breast Cancer Detection using Histopathological Images

Jitendra Maan [1], Harsh Maan [2]

[1] Head - AI and Cognitive Experience, Tata Consultancy Services Ltd - India
[2] DWH/BI Developer, Amdocs - India

**ABSTRACT**
Cancer is one of the most common and fatal diseases in the world. Breast cancer affects one in every eight women and one in every eight hundred men. Hence, our prime target should be early detection of cancer because the early detection of cancer can be helpful to cure cancer effectively. Therefore, we propose a saliency detection system with the help of advanced deep learning techniques, such that the machine will be taught to emulate actions of pathologists for localization of diagnostically pertinent regions. We study identification of five diagnostic categories of breast cancer by training a CNN (VGG16, ResNet architecture). We have used BreakHis dataset to train our model. We focus on both detection and classification of cancerous regions in histopathology images. The diagnostically relevant regions are 'salient'. The detection system will be available as an open source web application which can be used by pathologists and medical institutions.

## I. INTRODUCTION

Cancer is one of the most common diseases a living being can go through and the second major cause of death. It can generate in any part of the body and all of its types, the cancer cells divide and spread in the surrounding tissues. One of the most severe and most common types of cancer present in females is Breast cancer which could occur due to several reasons such as lifestyle, screening, and family history. Statistically, every 1 in 8 women and every 1 in 800 men have chances of being diagnosed with breast cancer. There are many ways to test and diagnose breast cancer which include a Breast exam, Mammogram, Ultrasound, taking a sample of breast tissues (Biopsy), and Breast Magnetic Resonance Imaging (MRI). The number of cases of Breast cancer is growing by the day in low and middle-income countries as compared to high-income countries, one of the major reasons for this is the expensive procedure to get yourself diagnosed along with the substandard treatment provided as compared to the high-income countries. Many of the resources that are required for the diagnosis of mammography are not available in LMICs. The mortality rate from breast cancer has not reduced with the diagnosis from clinical breast examination or self-examination due to poor or inadequate amount of resources. We are living in the age where digitization is ubiquitous in the field of diagnosis, yet pathology is still largely dependent on studies done with the help of microscopic evaluations of tissues on glass slides. Over a last few years, the prospect of Computer-assisted-diagnosis has increased due to the evolution of machine vision and digitally scanning microscopes due to which slides of tissue histopathology can now be saved in the form of a digital images. Accordingly, digital tissue histopathology has boosted image analysis and machine learning techniques.

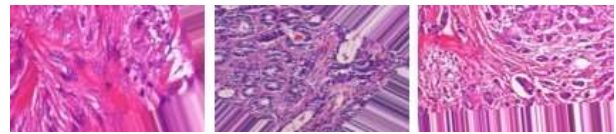

Fig. 1 Example of images available in BreaKHis Dataset.

The project focuses on both the detection and classification of cancerous regions in histopathological images. The diagnostically relevant regions are 'salient'. The motive can be clustered in two aspects: In theory, the classification task is studied manually on cancerous regions whereas a fully automated software allows automated slide image processing of the same. Detection and classifying only eliminate a small porting of computation and improves accuracy and efficiency. The prior detection step helps reduce false positives of classification without missing true positives. Detection of relevant regions is itself an important task that would lessen pathologists' workload significantly, where most of the cases are classified as benign or malignant. Further, the detection system would also assure that no critical region is overlooked during the diagnosis process.

## II. BACKGROUND STUDY AND RELATED WORKS

*A. Literature Survey*

There hasn't been a lot of work done in this field earlier. Though, recently, a lot of new approaches have come out. This is an emerging field which has a huge scope of improvement. Hence, our work aims at contributing significantly to the given field.

Few papers include Breast Cancer Diagnosis using Deep Learning Algorithm [1] which uses Wisconsin Breast Cancer Database. The mentioned database provides 569 rows and 30 features. They have used pre-processing algorithms like normalizer, label encoder etc. for scaled dataset. Their work was published in 2018. Another paper, titled, Breast Cancer Diagnosis from Histopathological Image based on Deep Learning uses BreakHis dataset. They have used CNN, convolutional neural networks to classify images as benign and malignant. But they have used Inception V3 model which is poor in initializing. A lot of computing time is wasted and it is expensive in making changes. Such work was published in 2019.

Another paper, titled, Nuclear Atypia Grading in Histopathological Images of Breast Cancer [2] Using Convolutional Neural Networks used CNN-based methodology. They used Notttingham's grading method which consists of 3 parts: Mitotic cell detection, nuclear atypia and tubule formation. Their model consists of feature extraction part which includes convolutional layers and activation function part which includes ReLu and pooling layers. The same was published in 2018.

Another method titled, Histopathological Image Analysis for Breast Cancer Detection Using Cubic SVM [3] which again uses BreakHis dataset. For experimental purpose they have tested six variants of SVM, that is, support vector machine along with random forest approach and KNN that is K Nearest Neighbors. They came to a conclusion that cubic SVM outperformed all the other techniques. But SVM struggles in predicting class labels when the size of class is large. This work was published in 2020. Yet another technique, titled, A Deep CNN Technique for Detection of Breast Cancer Using Histopathology Images [4] used BreakHis dataset. They also use CNN that is convolutional neural networks to classify images as malignant and benign. They use DenseNet which are extremely memory hungry networks. Their research was published in 2020.

A newer approach, titled, Breast Cancer Classification From Histopathological Images Using Patch-Based Deep Learning Modelling was published in 2021. In this paper, a new patch based deep learning method was introduced. It uses DBN, Deep Belief Network and then logistic regression is used to classify images. The accuracy of their model is just 86%. This is a new technique which is not very robust, and it was published in 2021.

Our model aims to be light on the machine and work efficiently in classifying images. We have used BreakHis dataset to classify images using CNN and have significantly improved on previous approaches.

*B. Dataset*

There are various data-sets available for histopathological images which can be used with deep learning methods for detection of tumor tissue. We have used BreaKHis dataset [6] which was collected by P&D Lab, Brazil during 2014 and composes 9109 microscopic images of breast tumor tissue of 82 patients using various magnifying factors (40X,100X,200X,400X). It is basically grouped into two categories, i.e, benign and malignant. Example of histopathological images in BreaKHis dataset are illustrated in Fig. 1. Benign indicates that tumor tissue is absent, whereas malignant indicates tumor tissue is present. The dataset consists of four malignant tumors: carcinoma, lobular carcinoma, mucinous carcinoma and papillary carcinoma. BreaKHis Dataset structure is illustrated by Table1.

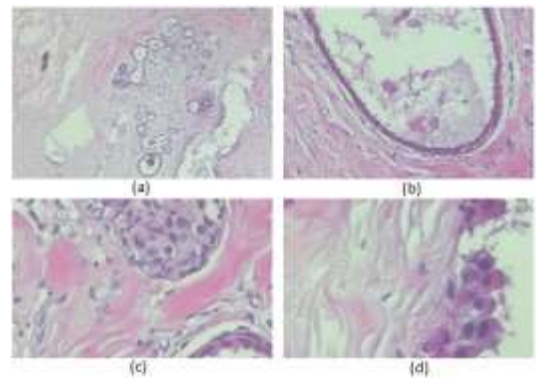

Fig. 2a. A slide of breast benign for the same patient seen in different magnifying factors: (a) 40X (b) 100X (c) 200X (d) 400X.

| Magnification | Benign | Malignant | Total |
|---|---|---|---|
| 40X | 652 | 1370 | 1995 |
| 100X | 644 | 1437 | 2081 |
| 200X | 623 | 1390 | 2013 |
| 400X | 588 | 1232 | 1820 |
| Total of Images | 2480 | 5429 | 7909 |

Table 1. BreaKHis Dataset 1.0

## III. METHODOLOGY

A practical solution to breast cancer detection problem involves feature extraction of histopathological images using CNN based method for detection of tumor cells. To this end, we present a deep learning CNN based methodology for detection of breast cancer using histopathology images. The Framework architecture is illustrated in Fig. 3. It can be divided into 2 parts based on its functionality, named feature extraction and activation function. Feature extraction includes convolutional network layers and activation function includes pooling layers, activation layers (ReLu, Softmax). We explain the key details about our method in the next subsections.

### A. Convolution Neural Network Architecture

Convolutional neural networks basically have layers of artificial neurons. Artificial neurons are nothing but mathematical functions which calculate weighted sum of multiple inputs and returns an activation value. When we insert an image into the convolutional network, several activation maps are generated by the layers. Neuron takes a patch of pixels as input, then they multiply it by the weight, add them and pass it through the activation function. This operation where we multiply pixel value by its weight and then adding them up is called as "convolution". Convolutional Layer. It is the most important mechanism for CNN model. The convolutional output layer is illustrated by the following equation:

$$I_{m_1,m_2} \star K_{k_1 \times k_2} = \sum_{i=(-k_1+1)/2}^{(k_1-1)/2} \sum_{j=(-k_2+1)/2}^{(k_2-1)/2} I_{m_1-i,m_2-j} \star K_{i \times j}.$$

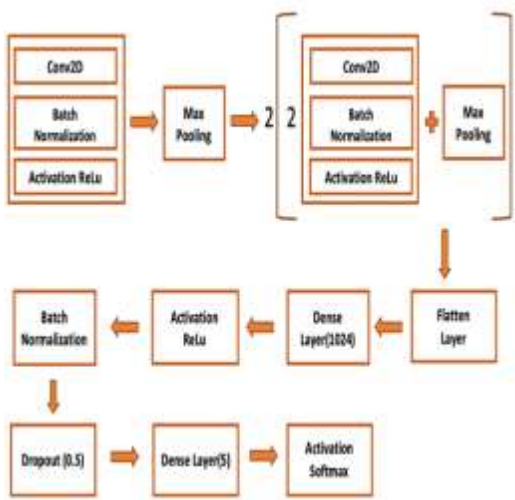

Fig. 3. Our Proposed framework architecture.

Neurons generates linear output. So when output of neuron is given to another neuron, it generates another liner output. To avoid and control this error, we introduce activation functions such as
- Sigmoid
- ReLU

Kernel, another important ingredient of convolutional layer basically scans the input data and extracts all the features. We have performed zero padding to conduct convolution operation perfectly. Sub sampling/pooling is performed to down-sample the features, i.e, to reduce the overall dimensionality and complexity. Four type of pooling operations are available. We have implemented Max-Pooling operation/layer in our CNNbased model.

Huge number of neurons are present in a deep neural network which allows the network to make large number of predictions. This basically enhances the performance of training data-set and reduces the performance of testing dataset. We have implemented Drop-out layer in our method to overcome this problem. In drop-out, a few neurons are randomly dropped to overcome the problem of over-fitting. Neurons are connected in a flattened way towards the end of the network. Deep convolution neural networks consist multiple processing layers which are learnt end to end jointly to address specific issues [9]-[11].

### B. Optimizers

Optimizers are nothing but algorithms which are used to change attributes of neural network such that weights and learning rate reduces. Several optimizers are available at the moment which have been researched in the last few couples of years and each of them have their own advantage and disadvantage. We tried 4 different optimizers with our CNN method model.

*1) Stochastic Gradient Descent (SGD)*

SGD let us update the network weights per each training image. Weights were updated based on the following equation:

$$w_{t+1} = w_t - \eta \frac{\partial C}{\partial w_t}$$

where following equation is gradient update equation and $\eta$ is the learning rate.

$$\frac{\partial C}{\partial w_t} = \nabla_w C\left(w_t; x^{(i)}; y^{(i)}\right)$$

We tried 3 different learning rates with SGD Optimizer and accuracies obtained with the same are illustrated by Fig. 4.

| Learning Rate (LR) | Testing Accuracy |
|---|---|
| 1e-2 | 89.72% |
| 1e-3 | 85.63 |
| 1e-4 | 82.25 |

Fig. 4. Learning rate v/s Accuracy for SGD Optimizer

*2) ADAM Optimizer*

Benefits of Nesterov momentum, AdaGrad, and RMSProp algorithms were combined together and made into ADAM optimizer algorithm. Weights are updated in the following equation:

$$w_t^i = w_{t-1}^i - \frac{\eta}{\sqrt{\hat{v}_t} + \epsilon} \cdot \hat{m}_t$$

where $\eta$ is the Learning rate. We tried 3 different learning rates with Adam Optimizer and accuracies obtained are illustrated by Fig. 5.

| Learning Rate (LR) | Testing Accuracy |
|---|---|
| 1e-2 | 85.39% |
| 1e-3 | 86.74% |
| 1e-4 | 89.41% |

Fig. 5. Learning rate v/s Accuracy for Adam Optimizer

*3) RMSprop Optimizer*

Problem where learning decreased monotonically led to the introduction of RMPprop algorithm.

$$G = \nabla_w C(w_t)$$
$$E[G^2]_t = \lambda E[G^2]_{t-1} + (1-\lambda)G_t^2$$
$$w_t^i = w_{t-1}^i - \frac{\eta}{\sqrt{E[G^2]_t + \epsilon}} \cdot \nabla_w C(w_t^i)$$

We tried 3 different learning rates with RMSprop Optimizer and accuracies obtained are illustrated by Fig. 6.

| Learning Rate (LR) | Testing Accuracy |
|---|---|
| 1e-2 | 86.59% |
| 1e-3 | 90.70% |
| 1e-4 | 90.45% |

Fig. 6. Learning rate v/s Accuracy for RMSprop Optimizer

*4) Nadam Optimizer*

Nadam algorithm is basically an extension of Adam algorithm combined with Nesterov momentum gradient descent.

$$w_t^i = w_{t-1}^i - \frac{\eta}{\sqrt{v_t + \epsilon}} \cdot \tilde{m}_t$$
$$\tilde{m}_t = \beta_1^{t+1}\hat{m}_t + (1-\beta_1^t)\hat{g}_t$$
$$\hat{g}_t = \frac{g_t}{1 - \prod_{i=1}^{t}\beta_1^i}$$

We tried 3 different learning rates with Nadam Optimizer and accuracies obtained are illustrated by Fig. 7.

| Learning Rate (LR) | Testing Accuracy |
|---|---|
| 1e-2 | 85.86% |
| 1e-3 | 85.15% |
| 1e-4 | 87.64% |

Fig. 7. Learning rate v/s Accuracy for Nadam Optimizer

## IV. EXPERIMENTAL RESULTS

We followed the protocol which is mentioned in [6], so basically the dataset is divided into training(70) and testing(30) sets considering patients in training set and testing set are unique and exclusive.

*A. Hyper-tuning*

Fine-tuning of parameters is essential in classification of histopathological images. Since neural network needs huge dataset or large number of images to be trained properly. However, medical images such as histopathological images are scant. Neural networks can learn connection between input and output automatically [13]. Some of these connections succeed during training period but do not with the testing datatset. This leads to over-fitting and performance degradation of the model [13]. This is why hyper-tuning of parameters is one of the most important task. Hyper-tuning of parameters helps us achieve best results with our model. Adam algorithm was used to optimize our proposed model [14,15]. We tried three different learning rates with 4 separate optimizers [15,16]. We used dropout 0.5 to prevent over-fitting of the model [17].

- Dropout: 0.5
- Epochs: 100
- Optimizer: Adam
- Batch size: 55
- Decay rate: ( lr/epochs)
- Learning rate: $10^{-3}$
- $2 \times 2$ max-pooling with stride 2
- First dense layer has 1024 dimensions
- Second dense layer has 5 for 5 classes
- Removed Dropouts in the convolution layers
- Reduce pool size for Better and finer patterns
- Increased the filters for the convolution layers (32-64-128)

*B. Image Augmentation*

Image augmentation is used to increase the dataset by making new modified images. By using the ImageDataGenerator in Keras library [12], we create new augmented histopathological images. With new augmented images, we ensure that our neural network will train better at every step and epoch. Images are firstly given to ImageDataGenerator, which converts and transforms every image by some random rotation, factor, etc. Parameters of augmented images are:

- Zoom range: 0.2
- Shear range: 0.2
- Rotation range: 42
- Width shift range: 0.2
- Height shift range: 0.2
- Horizontal flip: True
- Fill mode: Nearest

*C. Performance Analysis*

Performance Metrics for evaluating the efficiency of our proposed model are illustrated in Fig. 8. The training and validation accuracy curve of our model is illustrated in Fig. 9. Training and Validation loss curve of our model is illustrated in Fig. 10. The variation in validation accuracy compared to training accuracy can be reduced by increasing the dataset size. Similarly, the variation in validation loss compared to training loss can be reduced by increasing the number of histopathological images. Finally, we achieved training accuracy of 96.7% and testing accuracy of 90.4%.

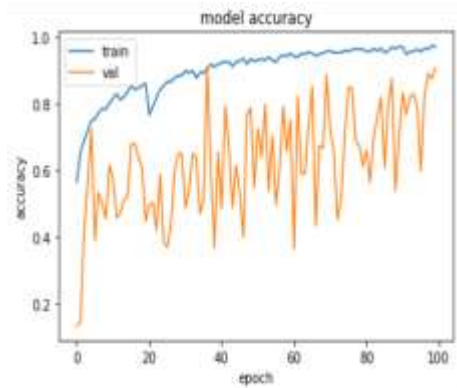

Fig. 8. Performance Metrics and their values illustrated in the table.

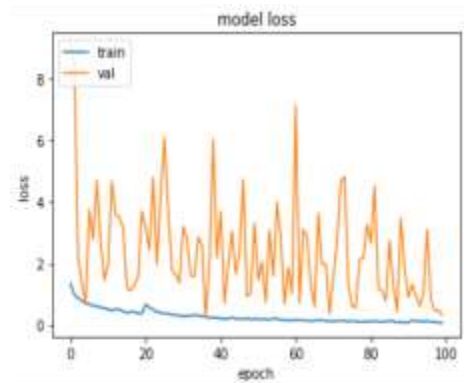

Fig. 9. The training and validation accuracy curves of fully trained and finetuned model.

Fig. 10. The training and validation loss curves of fully trained and fine-tuned model

## V. CONCLUSIONS AND FUTURE WORK

This report summarizes the work done in the field of medical science and technology. We have successfully detected malignant tissues in histopathological images using CNN. Our future work involves deploying a website where a user can put his/her details and get a result based on image input.